\documentstyle[12pt,epsf]{article}

\newcommand{\half}{{1\over2}}

\newcommand{\Tr}{{\rm Tr\,}}
\def\cN{{\cal N}}

\newcommand{\mysection}[1]{\setcounter{equation}{0}\section{#1}}

\newfont{\Bbb}{msbm10 scaled 1200}     
\newcommand{\mathbb}[1]{\mbox{\Bbb #1}}
\def\IR{{\mathbb R}}
\def\IZ{{\mathbb Z}}

\newcounter{fignum}
\newcommand{\figurnum}{\arabic{fignum}}
\newcommand{\figur}[2]{
\addtocounter{fignum}{1}
\addcontentsline{lof}{figure}{\protect
\numberline{\arabic{section}.\arabic{fignum}}{#2}}
\hspace{-4mm}{\it fig.}\ \figurnum.
\begin{figure}[t]\begin{center}
\leavevmode\hbox{\epsffile{#1.eps}}\\[3mm]
\parbox{10cm}{\small \bf Fig.\ \figurnum : \it #2}
\end{center} \end{figure}\hspace{-2mm}}
\newcommand{\be}{\begin{equation}}
\newcommand{\ee}{\end{equation}}
\newcommand{\eel}[1]{\label{#1}\end{equation}}
\newcommand{\bel}[1]{\begin{equation}\label{#1}}
\newcommand{\bea}{\begin{eqnarray}}
\newcommand{\eea}{\end{eqnarray}}
\newcommand{\eeal}[1]{\label{#1}\end{eqnarray}}
\newcommand{\baq}{\begin{equation}\begin{array}{rcl}}
\newcommand{\eaq}{\end{array}\end{equation}}
\newcommand{\eaql}[1]{\end{array}\label{#1}\end{equation}}
\newcommand{\beac}{\begin{equation}\begin{array}{rcl}}
\newcommand{\eeacn}[1]{\end{array}\label{#1}\end{equation}}
\newcommand{\ba}{\begin{array}}
\newcommand{\ea}{\end{array}}

\newcommand{\equ}[1]{(\ref{#1})}

\newcommand{\al}{{\alpha^{'}}}
\newcommand{\beq}{\begin{eqnarray}}
\newcommand{\eeq}{\end{eqnarray}}

\begin{document}
\begin{titlepage}
\newcommand{\preprint}[1]{\begin{table}[t]  
           \begin{flushright}               
           \begin{large}{#1}\end{large}     
           \end{flushright}                 
           \end{table}}                     
\begin{flushright}
NSF-ITP-00-30 \\
hep-th/0004131
\end{flushright}

\begin{center}
\LARGE{Sphalerons, Merons, and Unstable Branes in $AdS$}

\vspace{10mm}

\normalsize{
Nadav Drukker$^{1,2}$,
David J. Gross$^1$
and N. Itzhaki$^{3}$}

\vspace{10mm}

{\em $^{1}$Institute for Theoretical Physics,
University of California, Santa Barbara,
CA 93106}

{\em $^{2}$Physics Department, Princeton University,
Princeton NJ 08544}

{\em $^{3}$Physics Department, University of California, Santa Barbara,
CA 93106}

\vspace{5mm}
{\tt drukker, gross@itp.ucsb.edu, sunny@physics.ucsb.edu}

\end{center}

\vspace{10mm}

\begin{abstract}

We construct unstable classical solutions of Yang-Mills theories and their
dual unstable states of type IIB on $AdS_5$. An example is the unstable
D0-brane of type IIB located at the center of $AdS$. This has a field
theory dual which is a sphaleron in gauge theories on $S^3\times\IR$. We
argue that the two are dual because both are sphalerons associated to the
topology of the instanton/D-instanton. This agreement provides a
non-supersymmetric test of the $AdS$/CFT duality. As an illustration, many
aspects of Sen's hypothesis regarding the unstable branes can be seen easily
in the weakly coupled dual field theory description. In Euclidean $AdS$ the
D0-branes are dual to gauge theory merons. This implies that the two ends of
a D0-brane world-line carry half the charge of a D-instanton. Other examples
involve unstable strings in the Coulomb phase.

\end{abstract}

\end{titlepage}

\baselineskip 18pt

\mysection{Introduction}

Like any other strong/weak duality which cannot be proven directly, the
$AdS$/CFT duality \cite{Maldacena:1998re} was tested using BPS configurations.
Such configurations are protected by supersymmetry and can be traced while
interpolating from weak to strong coupling. Non-BPS configurations are
not protected and in general any result obtained using the duality is
considered to be a prediction rather then a test.

In this paper we study some non-BPS states of gauge theories at weak
and strong coupling. The configurations we discuss are unstable classical
solutions which sit at the top of non-contractible loops in configuration
space (sphalerons)
\cite{Taubes:1982ie, Manton:1983nd, Klinkhamer:1984di, Forgacs:1984yu}.

Let us remind the reader what a sphaleron is. Say there exists a one
parameter family of field configurations that form a non-contractible loop.
One should think of all homotopically equivalent loops
and find the point with maximal energy along each loop. Now consider the
minimum of all those energies, since the loops are not contractible, that
energy has to be greater than zero, and the corresponding field configuration
is a saddle point---the sphaleron. In practice, once one understands the
topology, it is usually easy to find the loop going through the sphaleron.
A schematic picture is given in
\addtocounter{fignum}{1}
{\it fig.}\ \figurnum.

If there is a d-dimensional topologically charged object in the theory,
then in general there would be a d+1-dimensional sphaleron. A simple example
is a theory which has an instanton. Then consider the one parameter family of
static field configurations where the extra parameter replaces the Euclidean
time. This family of field configurations has the same topological charge as
the instanton. By varying the parameter, one starts and ends at the vacuum,
and at the middle point there will be an unstable solution to the equations
of motion. It sits at the top of a non-contractible loop in the space of
field configurations. This is the sphaleron.

It was recently argued by Harvey, Ho\v rava and Kraus \cite{harvey} that
unstable D-branes of string theory
\cite{sen,Bergman:1998xv} are sphalerons. For example the type IIB D0-brane
can decay to the vacuum, but its existence is dictated by the same topology
as the D-instanton, whose charge is classified by K-theory
\cite{Witten:1998cd}. One can construct a one parameter family of static
configurations whose topology is that of the D-instanton. The D0-brane
sits at the top of the loop.
\addcontentsline{lof}{figure}{\protect 3}
\begin{figure}[t]\begin{center}
\leavevmode\hbox{\epsffile{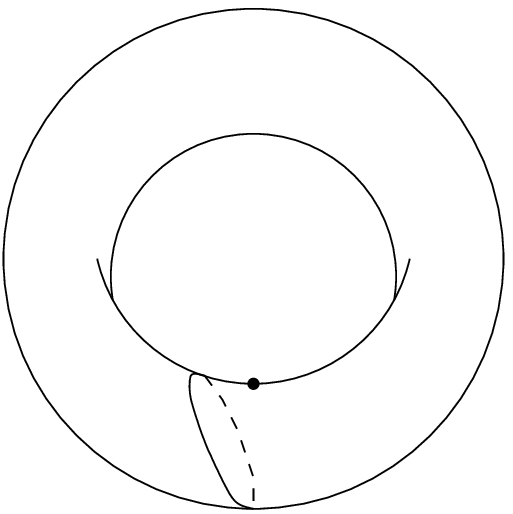}}\\[3mm]
\parbox{10cm}{\small \bf Fig.\ \figurnum : \it
The existence of a non-contractible loop, as illustrated in the picture
proves that there is an unstable saddle point. It is marked by the point
in the picture.}
\end{center} \end{figure}

This will serve as our first example. We consider the
configuration of a D0-brane at the center of $AdS$. This is
a massive, non-BPS object in the large $N$ and large coupling classical
limit of the theory. In global $AdS$ geometry, where the topology of the
boundary is $S^3\times\IR$, this is a static, spherically symmetric,
configuration.

A similar configuration exists at weak 't Hooft coupling. It is
explained in detail in Section 2, let us just say now that it is a ``half
pure gauge'' configuration. If one considers the $SU(2)$ instanton
\cite{Belavin:1975fg}, this is the configuration half-way through the
tunneling process, which is at the top of the potential. That is why it is
a solution of the equations of motion with one unstable mode.
This gauge theory sphaleron has many properties similar to the D0-brane
in $AdS$. It is static, spherically symmetric and has a single tachyonic
mode. We will argue that it is dual to the D0-brane in $AdS$.
We also find duals of the configuration with $k$ coincident D0-branes,
which have $k^2$ unstable modes, in string theory and in the gauge theory.

It is rather perplexing at first that we are able to find a dual description
for a non-BPS object. But there is, in fact a good reason for that. The
D0-brane sits in the middle of a non-contractible loop with the same
topology as the D-instanton, while the gauge theory solution is at the
middle of a loop with the topology of the gauge theory instanton which is
dual to the D-instantons.

Put differently, the instanton describes a
tunneling process under a potential barrier, and the sphaleron sits at the
top of the potential. The mass of the sphaleron is the maximum hight of
the potential. In the dual theory, the D-instanton also describes a tunneling
event, and the sphaleron is again at the top of the potential barrier. The
mass of the D0-brane is the hight of the potential. Since the YM instanton
and D-instanton are dual, they describe the same tunneling process in the
dual pictures. The shape of the potential is altered by quantum corrections,
but there is always an unstable point in the middle.

It is very simple to calculate the potential through which the instanton
tunnels, it is given by a quartic of the field. The potential of string
theory is much more complicated, understanding this potential is crucial to
proving the brane anti-brane annihilation procedure, which is in the heart
of Sen's construction, and the classification of D-brane charges by
K-theory. This issue was addressed recently by using level truncation in
string field theory \cite{sz} with impressive results. Our dual description
fits neatly with Sen's conjecture.

One should contrast this with other strong-weak dualities. It is more typical
for the topological excitations of one theory to become the elementary
excitations of the dual theory. For example the kink of the Sine-Gordon
model become the fermions in the dual Thirring model. The same is true in
S-duality of $\cN=4$ Yang-Mills (and type IIB), where the topologically
charged monopole goes over to the W-boson which is the elementary excitation.
Here we find that one topologically charged object goes to another
topologically charged object, and therefore there are sphalerons associated
to those topologies. Roughly speaking, the AdS/CFT duality is special since
it is a strong/weak duality with respect to the 't Hooft coupling, while
the solitons' masses are of the order of $1/g_{YM}^2$.

These ``half pure gauge'' configurations were considered in the past on
$\IR^4$. They are singular at the origin and at infinity, but the
singularities can be smoothed out. Those objects were named merons
\cite{merons}. The singularity at the origin and at infinity are replaced
with half an instanton, interpolating between the vacuum and the meron.

This has an exact analog in Euclidean $AdS$, where a D0-brane appearing out
of the vacuum, propagating and annihilating is dual to the meron. The
D0-brane follows a geodesic in $AdS$, and it's action depends logarithmically
on the separation of the two end points. The same
logarithmic behavior (up to a coefficient which depends on the 't Hooft
coupling) shows up on the gauge theory side. Because of the entropy of those
configurations, they might dominate the path integral for large $g_{YM}$.

We will also argue that each of the two end points of the D0-brane carries
half a unit of D-instanton charge. The D0-brane serves as a flux tube
carrying half a unit of flux from one end to the other, thus preserving the
Dirac quantization condition of D-instanton charge. A similar story applies
to higher dimensional branes, so the unstable D-branes can be regarded as
D-merons. Unlike $AdS$, where the action of the D0-brane is logarithmic,
in flat space it's linear, therefore it would not be dynamically favorable
for D-branes to break by this mechanism.

The paper is organized as follows.
We describe the details of the sphaleron on $S^3\times\IR$ and the D0-brane
in Lorentzian $AdS$ in section 2.
In Section 3 we describe the meron configurations. We review the
old construction in the gauge theory, and then we describe its dual. We
interpret the unstable branes as D-merons in Section 4.
In Section 5 we consider another example of a duality between unstable
classical solutions. We show that gauge
theories in the Coulomb phase admit unstable string solutions which do not
carry gauge invariant magnetic or electric fluxes. We describe the $AdS$
dual of this solution. The unstable string can also serve as a meron, and
we explain how a monopole can be separated into two halves as long as they
are connected by one of those strings.

\mysection{Sphaleron particle}

In this section we consider sphaleron particles in four dimensional $U(N)$
Yang-Mills theory, and their $AdS$ duals. Since Yang-Mills theory is a
conformal theory there are no static
finite energy (stable or unstable) solutions
on $\IR^4$ simply because there is no scale to fix the mass of the solution.
However, there is a sphaleron particle if we consider the gauge theory on
$S^3\times\IR$. In that case the size of the sphere, $R$, is the only
scale in the theory and so the mass of any static solution is $\sim 1/R$.

We consider first the perturbative YM description, and then the $AdS$ dual.
While the duality is true only for the theory with the $\cN=4$ matter content,
in perturbation theory the particle exists already in the pure gauge theory.

\subsection{Gauge theory description}

The topology that supports a stable particle in four dimensions is the map
from the $S^2$ at spatial infinity to the fields. For $U(N)$ pure gauge
theory the only relevant topology is $\pi_2 (U(N))=0$. Hence this theory
does not admit any topologically charged stable particles (on either
$\IR^4$ or $\IR\times S^3$). However, since
\bel{homo}
\pi_{2l+1}(U(N))=\IZ\,,
\qquad\mbox{for}\qquad
l<N\,,
\ee
there are unstable solutions to YM theory. These solutions, which we describe
below, sit at the top of a non-contractible $S^{2l-1}$ in configuration space.

We start by considering the simplest case of $l=1$. In that case we have
a non-contractible loop  in the configuration space of $SU(2)$ gauge theory
which we embed in $SU(N)$. The topology of the non-contractible loop is the
same as the instanton topology. It is useful to recall the instanton
solution, it is given by the ansatz
\be
A_\mu=-if(r)\partial_\mu U U^\dagger\,,
\qquad
U={x^\mu\sigma_\mu\over r}={x_0+ix_i\sigma_i\over r}\,,
\qquad
r^2=x_0^2+x_i^2\,,
\eel{inst}
where $\sigma_i$ are the Pauli matrices and $x_0,x_i$ the four Euclidean
directions.
The Yang-Mills action now yields
\be
S={1\over 4g_{YM}^2}\int_0^\infty dr\, 96\pi^2
\left({r\over 2} f^{\prime 2}+{2\over r}f^2(1-f)^2\right).
\eel{r4a}
The equations of motions have three constant solutions $f=0$, $f=1$ and
$f=1/2$. $f=0, 1$ are  stable solutions which correspond to two vacua.
The instanton solution, $f(r)=r^2/(a^2+r^2)$, interpolates between $f=0$ at
the origin and $f=1$ at infinity. The configuration with $f=1/2$ is an
unstable solution, it solves the second order equation of motion, but
unlike the two vacua and the instanton solution, does not solve the first
order BPS equation.

On $\IR^4$ we see from (\ref{inst}) that $f=1/2$ is a non-static singular
solution. It was first discussed in \cite{deAlfaro:1976qz} and was studied
further in \cite{merons}. Those are the merons which we will discuss in the
next section. On $S^3\times\IR$ however, the solution is static, regular
and completely delocalized\footnote{
Since the solution is smeared over the entire $S^3$, it could be considered
a tachyonic vacuum, rather than an unstable particle. Since the space is
compact, it is hard to distinguish between the two notions.}
on $S^3$. To see this, note that the conformal transformation that takes
$\IR^4$ with metric $ds^2=dr^2+r^2d\Omega_3^2$ to $S^3\times\IR$ with metric
$ds^2=dt^2+R^2d\Omega_3^2$ is
\bel{exp}
r=\exp(t/R).
\ee
Therefore the action of the sphaleron on  $S^3\times\IR$ is
\be
S=\int_0^\infty dr\,{3\pi^2\over g_{YM}^2r}=
{3\pi^2\over g_{YM}^2R}\int_{-\infty}^\infty dt\,.
\eel{acsp}
We see that the action does not depend on $t$ and that the sphaleron mass is
\be
M_{Sp}={3\pi^2\over g_{YM}^2R}.
\ee

A non-contractible loop of static field configurations going between the
two vacua and through the sphaleron is given by \equ{inst} with
\bel{ncl}
f(r)=\alpha,
\qquad
0\leq \alpha\leq 1.
\ee

Equation (\ref{inst}) implies that for constant $f$ we get $A_r=0$ (on
$\IR^4$) and hence $A_t=0$ (on $S^3\times\IR$) and that $A_{\theta}$ does not
depend on $t$. Therefore, $F_{t\theta}=0$ (where $\theta$ represents the
$S^3$ coordinates). This has important implications for the non-contractible
loop. First, the field configurations along the entire non-contractible
loop (\ref{ncl}) do not depend on $t$, and can be described in terms of
the three dimensional theory on $S^3$.
Second, even though the conformal map with Lorentzian signature (see e.g.
\cite{Horowitz:1999gf}) is different from the Euclidean conformal map
(\ref{exp}), Wick rotation to Lorentzian signature (on $S^3\times\IR$) is
trivial along the entire non-contractible loop. This is not the case for the
instanton solution, which depends on $r$. Finally,
\bel{fft}
\Tr F\tilde F=0\,,
\qquad\hbox{while}\qquad
\Tr F^2=\frac{6}{R^4}\neq 0\,.
\ee
These features will prove to be important for the dual description, as we
shall see in the next section.

Next we turn to the cases when $l>1$. In those cases the solution exists
only for $SU(N)$ with $N>2$. Finding all sphaleron solutions for $SU(N)$
gauge theory is beyond the scope of the paper. However, there is a very
simple construction which yields sphalerons related to arbitrarily high
homotopy groups. Those are dual to the coincident D0-branes in $AdS$.

We can generalize the spherically symmetric ansatz \equ{inst} to larger
gauge groups by replacing the Pauli matrices and the identity by
\bel{a2}
A_\mu=-if(r)\partial_\mu U U^\dagger\,,
\qquad
U={x^\mu\gamma_\mu\over r}\,,
\ee
where the $\gamma$'s satisfy the algebra
$\gamma_\mu\gamma_\nu^\dagger+\gamma_\nu\gamma_\mu^\dagger=2\delta_{\mu\nu}$.
We use the simple choice
\be
\gamma_\mu
=\sigma_\mu\otimes I_k
=\pmatrix{
\sigma_\mu&0         &\cdots&0        \cr
0         &\sigma_\mu&\cdots&0        \cr
\vdots    &\vdots    &\ddots&\vdots   \cr
0         &0         &\cdots&\sigma_\mu},
\eel{matrix}
where $I_k$ is the identity matrix of rank $k$.
It is easy to see that this is still a solution of the equations of motion
if $f=1/2$. The action simply scales as the rank, $2k$,  of the matrices
$\gamma_\mu$. Therefore the mass of the k-sphaleron is
\be
M_k=k M_{Sp}\,.
\ee

This sphaleron solution has $k^2$ unstable modes, which correspond to each of
the $2\times2$ entries in the matrix in \equ{matrix}. The number of unstable
modes alone does not fix  the topology of the non-contractible loops
associated with the sphaleron. For example, the fact that we have $k^2$
unstable modes does not mean that the sphaleron sits at the top of $S^{k^2}$.
This would be inconsistent with $\pi_{k^2} (U(N))=0$ for even $k$. In fact
the topology is exactly that of $U(k)$. The sphaleron sits at the point
$-I_k$ in the group, which is opposite to the identity\footnote{
We described the sphaleron as the point in the algebra of $U(k)$ with
$f=\half I_k$. In the group that corresponds to $\exp(2\pi if)=-I_k$.}.
The $k^2$ unstable modes are the tangent vectors
in the algebra of $U(k)$. Since the group $U(k)$ has non-contractible
$S^{2l-1}$ for all $0< l\leq k$, there are such loops going through the
sphaleron. So we can choose to classify the tangent directions by those
spheres. All together there are indeed
$1+3+\cdots+2k-1=k^2$ unstable directions. The sphaleron sits, therefore,
at the top of $S^1,S^3,\cdots,S^{2k-1}$. In the next
section we shall see that this fit neatly with the results of \cite{harvey}.

Let us show this explicitly for $k=2$. Consider
\be
A_\mu=-i\partial_\mu U U^\dagger\otimes H\,,
\ee
where $U$ is of rank two, as defined in \equ{inst}, and $H$ any $2\times 2$
Hermitian matrix. We can parameterize
\be
H=\half(1+\alpha) I_2+\half\beta_i\sigma_i\,.
\ee
The sphaleron is at $\alpha=\beta_i=0$, which has $H=\half I_2$. Two vacua
are given by $\alpha=\pm1$, $\beta_i=0$, so that $H=0,1$. There is another
family of vacua, at $\alpha=0$, $|\beta|=1$, those are parameterized by an
$S^2$, the direction of $\beta_i$. Those vacua give $H$ with one eigenvalue
equal to zero and the other equal to one.

Identifying the two vacua at the end of the interval $-1\leq\alpha\leq1$
gives the non-contractible $S^1$. The parameters $\beta_i$ (with
$|\beta|<1$) take values in the ball $B^3$. Identifying all the boundary
points gives a non-contractible $S^3$.

The parameter $\alpha$ in \equ{ncl} gives a one-dimensional family of
configurations in $SU(2)$. In the previous paragraphs $\alpha$ and
$\beta_i$ gave a one and a three dimensional family of configurations
in $SU(4)$. Those are actually related to the non-trivial $\pi_3$ of
$SU(2)$ and to the non-trivial $\pi_3$ and $\pi_5$ of $SU(4)$. This is
true in general. To see this we have to include the spatial manifold
$S^3$.

The parameters $\alpha$, $\beta_i$ and the higher dimensional ones
live in $B^{2l-1}$. At every point there is a static field configuration
on $S^3$. So we have an $S^3$ for every point in $B^{2l-1}$. At
the boundary of the ball the field configuration is the vacuum, which is
trivial on the $S^3$, so we can take to sphere to shrink to a point. This
fibration of $S^3$ over $B^{2l-1}$ gives $S^{2l+2}$. Now recall the
well known fact that if the gauge group has a non trivial $\pi_{2l+1}$
then there is a non-trivial gauge bundle over $S^{2l+2}$ (the map from
$S^{2l+1}$ to the group is the transition function on the equator of
$S^{2l+2}$).

In the simplest case, adding the parameter $\alpha$ to $S^3$ allows us
to build an $S^4$, on which there are configurations with the topology
of the instanton.

\subsection{Supergravity side---Unstable D0-branes in $AdS_5\times S^5$}

The $AdS$/CFT duality is a strong/weak duality and
as such it takes classical configuration of one description into a quantum
excitation of the other description. Therefore, it is very hard to trace a
generic (non-BPS)  classical solution of weakly coupled SYM to the $AdS$
description. A sphaleron is a non-supersymmetric solution sitting at the
top of a non-contractible loop in the {\em classical} configuration space.
Therefore, it is natural to suspect that the quantum corrections will blur
the non-contractible loop. And that by the time the 't Hooft coupling is
large there will be no trace of the non-contractible loop and the sphaleron.

However, as we saw, the non-contractible loop associated with the sphaleron
of the previous subsection is described by the topology of the instanton. The
dual of the instanton is a D-instanton in $AdS$, which carries a charge in
K-theory. And so we should look for a non-contractible loop with
the topology of the D-instanton. Such non-contractible loops in flat
space-time were constructed in \cite{harvey}. There it was argued that the
sphaleron at the top of the loop is the type IIB D0-brane. We claim,
therefore, that the dual of the solution of the previous section are the
unstable D0-branes located at the origin of $AdS$. This is illustrated in
\figur{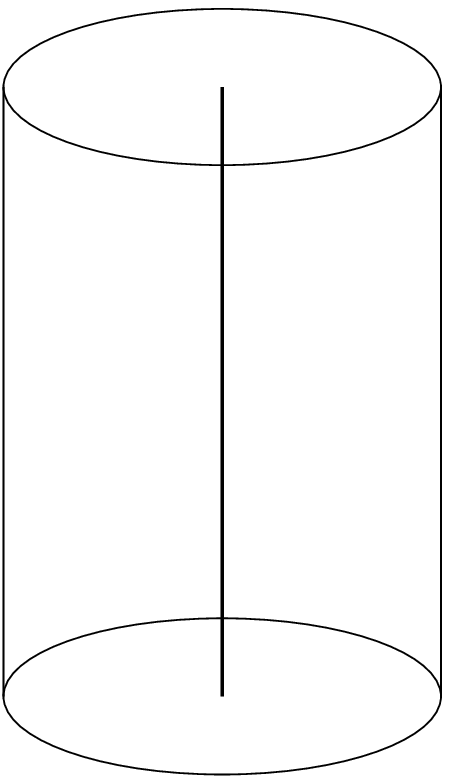}{An unstable D0-brane in the center of $AdS_5$.
The vertical direction is time, and the radial direction is the radial
coordinate of $AdS$. The boundary of global $AdS_5$ is
$S^3\times\IR$.}

Let us mention a few  properties of the unstable D0-branes and how they fit
into the claim that they are dual to the field theory  sphalerons.

\begin{itemize}

\item
A D0-brane (or $k$ coincident D0-branes) which are located at the center
of $AdS$  are static objects with respect to the global time.
Therefore they correspond to static objects in the gauge theory. The center
of $AdS$ corresponds to the extreme infra-red of the gauge theory, so the
energy is uniformly distributed over $S^3$.

\item
 From the closed string theory point of view the low energy supergravity
fields which are excited by the D0-branes are the NS-NS graviton and dilaton.
The RR-fields are not  excited. Using the dictionary of \cite{gkp,w} that
would correspond to $\Tr F\tilde F=0$ and to $\Tr F^2\neq0$, in agreement
with the field theory results (\ref{fft}). Note that the mass of the D0-brane
(and $\Tr F^2$) do receive quantum corrections for they are not protected by
supersymmetry,\footnote{The origin of the $\sqrt{2}$ is the fact that
the open strings living on an unstable brane carry two Chan Paton factors
$I$ and $\sigma_1$ \cite{Sen:1999mg}.}
\be
M_{D0}={\sqrt2 \over g_s\sqrt{\alpha'}}
={4\sqrt2\pi\lambda^{1/4}\over g_{YM}^2R}\,.
\eel{d0mass}

\item
In \cite{harvey} it was shown that the type IIB D0-branes are sphalerons of
string theory. That is, in flat space-time they sit at the top of a
non-contractible loop in the configuration space of string theory. Since for
large 't Hooft coupling the ``center'' of $AdS$ can be approximated by flat
space-time, one can simply embed the construction of \cite{harvey} in $AdS$.
There is also a global way to construct the D0-branes in $AdS$. Starting with
a system of D1-brane anti D1-brane stretching all the way to the boundary of
$AdS$, just like in flat space-time this system contains a complex tachyon
mode which can support an unstable D0-brane.

\item
It was further argued in \cite{harvey} that $k$ coincident D0-branes, which
have $k^2$ tachyonic modes correspond to sphalerons at the top of
$S^1,S^3,\cdots,S^{2k-1}$ in $U(k)$. This is exactly what we found from the
field theory side. It is worth while to note that in both descriptions the
mass is proportional to $k$.

\item
The NS sector of the excitations living on the D0-branes contains a real
scalar tachyonic mode. According to Sen's conjecture at the bottom of the
tachyon potential the negative energy cancels the tension of the brane and
we are left with the vacuum. This was tested, to a good accuracy, via level
truncation method in string field theory \cite{sz,bsz,wati}.
On the field
theory side we see that indeed the bottom of the potential ($f=0,~1$ in
(\ref{r4a})) is the vacuum. While calculating the tachyon potential in
string theory is complicated, in the field theory it's just a quartic
(\ref{r4a}).

\item
Since the tachyon is real, the potential can support a stable lower
dimensional brane. A  D-instantons in our case. Again, the energy of such a
configuration was calculated in string field theory with impressive
agreement with expectations \cite{bsz}.
On the field theory side the instanton indeed interpolates
between the two minima of the potential.

\item
Of all the instanton solutions on $\IR^4$, the one of radius $R$ centered
around the origin is special when translating to $S^3\times\IR$. It goes over
to a spherically symmetric solution on $S^3$. In that theory, this instanton
can be described as a quantum mechanical tunneling process between the two
minima of the quartic potential in (\ref{r4a}). The gauge theory sphaleron
sits at the middle of the potential. The width of the potential is $R$ and the
hight, which is the mass of the sphaleron, is proportional to $1/g_{YM}^2R$.
The action of the instanton is the area under the potential. In string
theory the same is true, only that $R$ is replaced by $l_s$. The hight
of the potential $\lambda^{1/4}/g_{YM}^2R=1/g_{YM}^2l_s$, and the width is of
order $l_s$. Since the action of the D-instanton is the same as the gauge
theory instanton, the area is the same, but the shape is altered.

\end{itemize}

We see therefore, that indeed the field
theory sphaleron is dual to the unstable D0-branes in $AdS$. It is important to
emphasize \cite{harvey} that the D0-branes are {\em not} sphalerons of
the low energy supergravity. That is, there is no supergravity
solution associated with the non-BPS D-branes which sits at the top of
a non-contractible loop of field configurations of the classical
supergravity.  The unstable branes are sphalerons of the full string
theory including all the quantum corrections to the sigma model.  Since
the full string theory on $AdS$ contains all the information about the
dual SYM theory it is not surprising that {\em in principle}
the field theory sphalerons
can be described by string theory on $AdS$. What is remarkable is that
the description is so simple.

A natural question that arises is whether the dual weakly coupled description
sheds new light on the diagonal U(1) problem associated with the unstable
D0-branes. Unfortunately, even though we can trace the D0-branes to the
weakly coupled region, we cannot trace the gauge theory living on them to the
weakly coupled description. Thus, as far as we can tell, the dual
description does not lead to any new insight on the U(1) problem. It is
worth mentioning that this problem of tracing the gauge theory
living on the brane to the weakly coupled description is not special to
D0-branes. For example,  we know that the dual of a D1-brane stretched
all the way to the boundary is the BPS monopole. But in weakly coupled field
theory there are no fields living on the monopole, while there is a $1+1$
gauge theory living on D1-branes in $AdS$. The reason is that
the size of the D1-brane is larger than the string scale only for
large 't Hooft coupling and so for small coupling the excitations which
were supposed to live on the monopole  cannot be separated  from the other
excitations.

It is interesting to note that when we have $k$ D0-branes the full topology of
the non-contractible loop, $U(k)$, with its non-contractible
$S^1,S^3,\cdots,S^{2k-1}$,
can be interpolated from the weakly to the strongly coupled region.
The $S^1$ is ``protected'' by the instanton which is BPS.
It should be interesting to understand why the other spheres are ``protected''
as well.

We would like to end this section with a comment on finite $N$. Our
construction of the field theory solution which is dual to k coincident
D0-branes is valid for $k\leq N/2$. Equation (\ref{homo}) implies that a
dual solution should be found at up to $k=N$. Presumably, a more
complicated ansatz will indeed yield the right solution. It should be
interesting to see if the mass is still linear with $k$. Another question
is what happens when $k>N$. In the field theory side we get out of the
stable regime. Is there any stringy exclusion principle associated with that?
Recall that the global construction of  $k$ D0-branes in $AdS$ involves
$k$ D1-branes and anti-D1-branes stretched all the way to the boundary
(this is a simple generalization of the discussion in \cite{harvey}).
Now, when $k=N$ the D1-branes can end on a NS-brane which wraps $S^5$
\cite{witten,go}. So it seems that the existence of a baryon vertex in
$AdS$ is the underlying mechanism which bounds the number of coincident
D0-branes in $AdS$ to $N$. Clearly, it would be nice to understand this
better.

\mysection{Merons in gauge theories and in $AdS$}

In Section 2.1 we studied the field configuration of ``half pure gauge''
on $S^3\times\IR$, and interpreted it as a sphaleron. As we mentioned, those
same configurations can be considered in the Euclidean theory on $\IR^4$,
they are still classical solutions, but there is a singularity at the origin
and at infinity. By smoothing out the singularities one gets a configuration
that solves the equations of motion almost everywhere and has finite action.
Those are the merons \cite{merons}.

We give a brief review of the merons in gauge theories and then will find
analogous configurations in string theory on $AdS$.

\subsection{Short review of merons}

Let us write again the instanton ansatz (\ref{inst})
\be
A_\mu=-if(r)\partial_\mu U U^\dagger\,,
\qquad
U={x^\mu\sigma_\mu\over r}={x_0+ix_i\sigma_i\over r}\,,
\qquad
r^2=x_0^2+x_i^2\,.
\ee
$f=0,1$ are vacuum solutions, and $f=\half$, the meron, is an unstable
solution which is singular at $r=0,\infty$. The action (\ref{acsp}) is
logarithmically divergent
\be
S={3\pi^2\over g_{YM}^2}\int_0^\infty {dr\over r}\,.
\ee
To regularize this divergence consider the following configuration
\be
f(r)=\left\{
\matrix{
\displaystyle{r^2\over r^2+R_1^2}\,, && r<R_1     \cr\cr
\displaystyle{1\over2}\,,            && R_1<r<R_2 \cr\cr
\displaystyle{r^2\over r^2+R_2^2}\,, && R_2<r\,.  }
\right.
\eel{halfhalf}
This is the meron for $R_1<r<R_2$, glued to half an instanton at the origin
and half at infinity. This carries the same topological charge as the
instanton, but it is broken in two parts. If one takes $R_1=R_2$, the
instanton solution is recovered. For $R_1\neq R_2$ this is a solution of the
equations of motion everywhere but at the spheres which separate the three
regions.

This is illustrated in
\figur{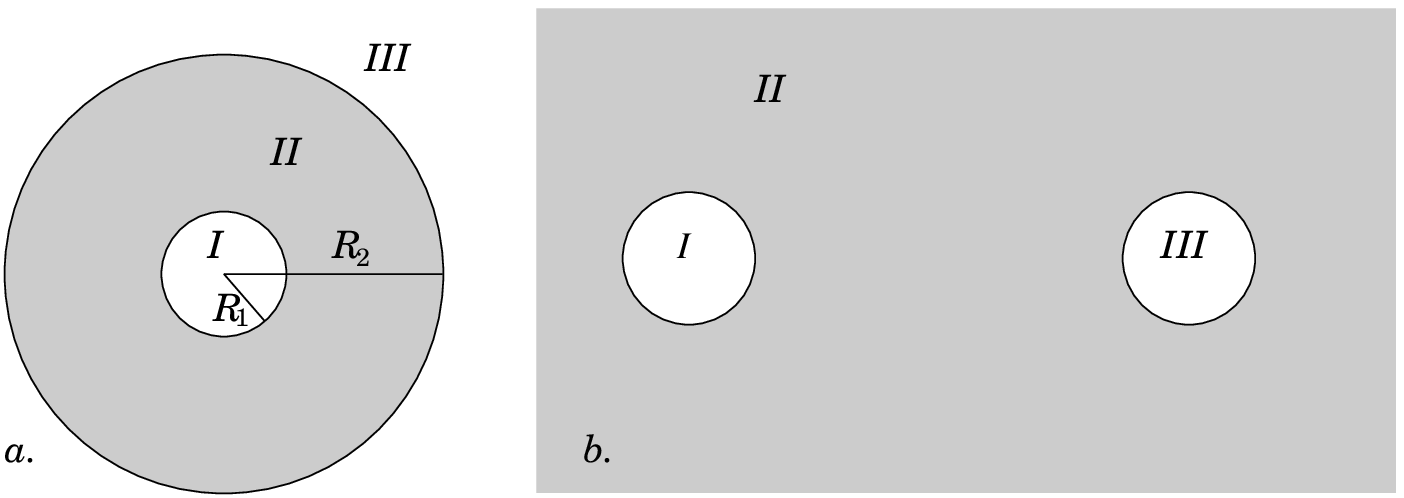}{a. The meron configuration. Region I is half an instanton, region
II is the meron with exactly half a pure gauge transformation, and region
III is another half instanton. By a large conformal transformation that takes
the point at infinity to finite distance and region III to a finite sphere
this can be mapped to the two meron configuration b.} {\it a.}
Region I and III are the half instantons near the origin and infinity. Region
II is the meron which connects the two. The action can be easily calculated,
and is equal to
\be
S=
{8\pi^2\over g_{YM}^2}
+{3\pi^2\over g_{YM}^2}\ln{R_2\over R_1}\,.
\eel{merac}

Since classical YM is conformally invariant, we can use a large gauge
transformation to map region III to a sphere at finite distance. The new
configuration is shown in {\it fig 3. b.}
Region I and III each carry half the topological charge of the instanton,
so at infinity this configuration is pure gauge.

One can, of course, replace the meron with an anti-meron, where instead of
half an instanton there is half an anti-instanton. The meron anti-meron pair
will have zero topological charge and two anti-merons $-1$ topological
charge. The interaction between a meron and and anti-meron is the same as
that between two merons.

The action of a  meron grows with the distance. Thus a first guess is that
the contribution of merons to the partition function is negligible. However,
the action grows only logarithmically so it can be compensated by a
large entropic factor.\footnote{
In thermodynamics this is, of course, common. At finite temperature one
has to minimize the free energy, $F=E-ST$ rather then the energy. Thus
a phase transition between minimizing $E$ and maximizing the entropy can
take place. Here the coupling constant plays the role of the temperature.}
The entropy contribution to the partition function goes like $L^4$, hence
the partition function associated with a meron is
\be
Z\sim L^4 \exp\left(-{1\over g_{YM}^2}\ln L\right)
= L^{(4-1/g_{YM}^2)}\,.
\ee
This suggest a phase transition at $g_{YM}^2\geq{1\over 4}$, wherein the
meron charges that made up the instanton dipole are liberated.
In the non-supersymmetric theories it was suggested that the appearance of
this new phase at large coupling, or large scale size, is closely related
to confinement, where the merons play the role of the three dimensional
instantons in Polyakov's mechanism for confinement \cite{Polyakov:1977fu}.
However, the full story is much
more complicated for one has to consider a gas of merons and their
interactions. This, as well as the fact the coupling runs, made it
very hard to estimate the relevance of merons to confinement.

Even though the coupling does not run for $\cN=4$, the main problem of
understanding the interactions among the merons is still very complicated.
In fact, in the $\cN=4$ theory, because of the fermions and scalars and the
fact that a meron breaks all supersymmetry, it is probably even more
complicated. We however cannot resist the temptation of speculating that
meron physics might be a clue for understanding $\cN=4$ theory at the
self-dual point ($g_{YM}^2=2\pi$).

\subsection{Merons in $AdS$}

We would now like to describe merons in the strong coupling limit of the
field theory, using string theory on $AdS$. We saw in Section
2 that the sphaleron solution of the gauge theory on $S^3\times\IR$ is
described in the dual theory by an unstable D0-brane. Since the meron is the
same field configuration as the sphaleron, only on $\IR^4$, it is also
described by a D0-brane in Euclidean $AdS$. Here we use the metric
\be
{ds^2\over \alpha'}
={\sqrt\lambda \over U^2}dU^2
+{U^2\over\sqrt\lambda} dx^2.
\ee

Consider a D0-brane which is created at some point $U_1$, propagates till
$U_2$ (and the same point in $\IR^4$) and annihilates. This is the $AdS$
dual of the configuration (\ref{halfhalf}) which was illustrated in
{\it fig. 3a.} By the UV/IR relation, for $U_1>U_2$, the internal circle has
a radius $R_1=\sqrt\lambda/U_1$ and the external circle
$R_2=\sqrt\lambda/U_2$.

The action of this configuration is
\be
S=S_{{\rm D}(-1)}+ T_{{\rm D}0}\int ds
={2\pi\over g_s}+{\sqrt2\lambda^{1/4}\over g_s}\ln{U_1\over U_2}\,,
\ee
where the first term ${2\pi/g_s}=8\pi^2/g_{YM}^2$ is equal to the instanton
action and is related to the creation of the brane and its annihilation, like
in the gauge theory. This contribution will be justified in the next section.
Comparing this to the gauge theory result (\ref{merac}), the constant part of
the action is unchanged, but the coefficient of the log is renormalized by
a factor proportional to $\lambda^{1/4}$, like the sphaleron mass
(\ref{d0mass}). Again, one should not be surprised, since this is a non-BPS
configuration.

Just as was explained in the previous section a conformal transformation
will take this geodesic into a D0-brane which is  created and annihilated
at the same value of $U$, but at a distance $L$ on
$\IR^4$, this is the $AdS$ dual of the configuration in {\it fig. 3b.}
The size of the two half instantons is simply $R=\sqrt\lambda/U$
Those two configurations are shown in
\figur{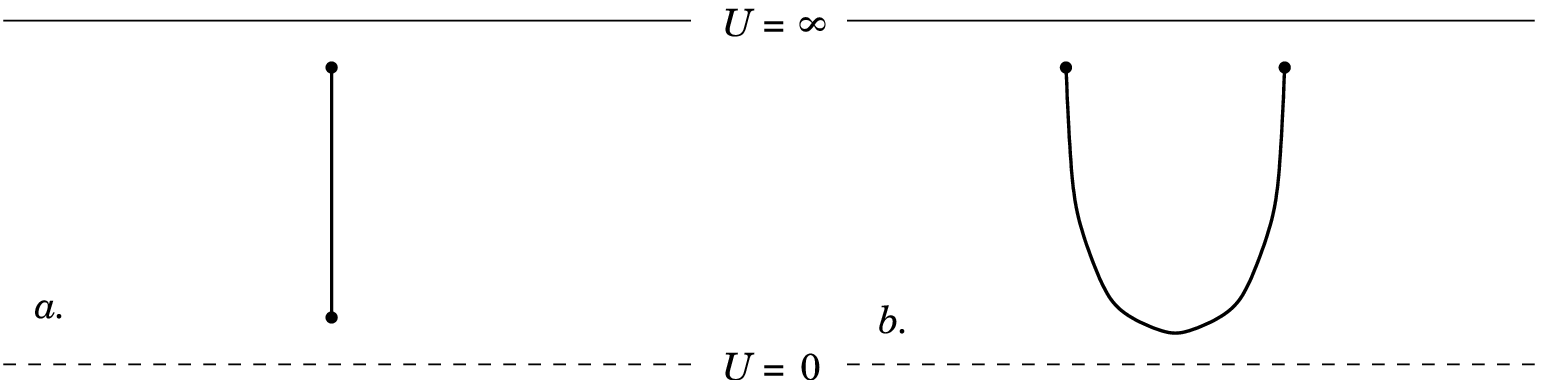}{
Two examples of unstable D0-branes created and annihilated in Euclidean
$AdS$. The boundary of $AdS$ is marked by the solid line at $U=\infty$.
Between the creation and annihilation point the particle travels along a
geodesic.}
It is not surprising, therefore, that the corresponding action is
\be
S=\frac{8\pi^2}{g_{YM}^2}+\frac{4\pi\sqrt{2}\lambda^{1/4}}{g_{YM}^2}\ln(L/R).
\ee

The fact that the logarithmic term is now proportional to
$\lambda^{1/4}/g_{YM}^2$, rather then just $1/g_{YM}^2$ as in the weakly
coupled theory seems to imply that the entropy contribution cannot compete
with the energy in strong coupling. That is,
\be\
Z\sim L^4 \exp \left(-{\lambda^{1/4}\over g_s} \ln(L)\right)\,.
\ee
So a phase transition at $g_s\sim 1$ is very unlikely for large $\lambda$.

\mysection{Unstable branes as D-merons}

In the previous section we studied D0-branes in Euclidean $AdS$. Since they
are unstable they can appear out of the vacuum, propagate some distance and
disappear again. This was dual to the meron in the gauge theory which
connects two regions where there are half instantons. Since the $AdS$ dual
of the instanton is the D-instanton, it is natural to suspect that at
each end of the D0-brane sits half a D-instanton.

We reached that conclusion by studying D0-branes in $AdS$, but this is true
in any string theory background, and the argument does not have to rely on
the $AdS$/CFT correspondence. After all, the D0-brane is a sphaleron at the
top of a non-contractible loop with the same topology of the D-instanton.
Therefore the entire event of a D0-brane creation, propagation and
annihilation can carry a unit of D-instanton charge. In fact, it can carry
1, 0, or $-1$ units of D-instanton charge.

The creation or annihilation of a D0-brane is an event that carries half (or
minus a half) of D-instanton charge. This might seem to contradict the charge
quantization condition. The product of the charge of a single D7-brane and
the charge of a single D-instanton is $2\pi$, so how can a D-instanton break
in two? The answer is that the two halves of the D-instanton are connected
by a D0-brane, which must carry half a unit of D-instanton flux.

This is analogous to a bar magnet, or a solenoid in electro-magnetism.
Outside the magnet the magnetic field looks like that of two separated,
oppositely charged, monopoles. But the monopole charge need not satisfy the
Dirac quantization condition, as the magnet (or solenoid), carries the flux
from one to the other.

It is amusing to push this analogy further. Just as the magnetic field in a
magnet is created by the angular momentum of the electric charges, the
D0-brane can be regarded as a very thin solenoid in which a current of
D7-brane charge produces a dual flux, connecting the one-half D(-1) charges.
It would be interesting to pursue this analogy further.

Since the unstable D0-branes connect pairs of $1/2$ D-instantons, they could
be called D-merons.

Thus far we considered only D0-branes, but the same is true for higher
dimensional branes as well. A D1-brane can break into two halves with an
unstable D2-brane in the middle. That is the same as saying that the boundary
of a Euclidean D2-brane could carry half-D1-brane charge.
Likewise in type IIA, a D0-brane can break in two with an unstable D1-brane
in the middle, and so on. A D2-brane ending on two half D1-branes is shown
in
\figur{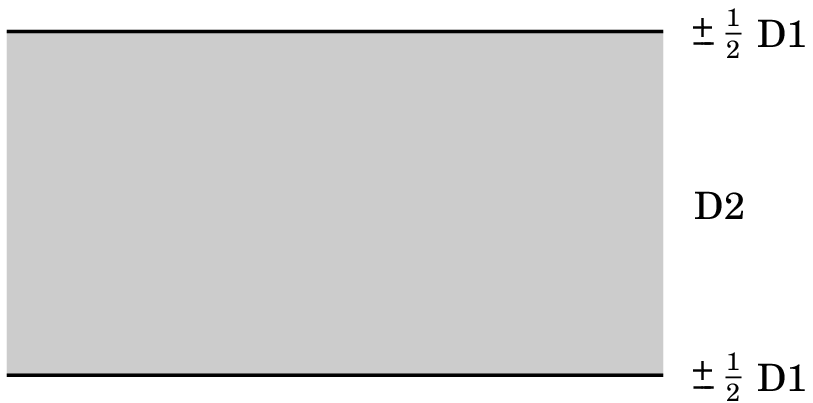}{
An unstable D2 brane of type IIB can end on two half D1-branes.}

In $AdS$ the action of the D0-brane is logarithmic, however in flat space it
will be linear. Therefore half D-instantons are clearly confined in flat
space. The same is true for the higher dimensional half-branes.

\mysection{Unstable strings in the Coulomb phase}

In previous sections we discussed how the existence of the instanton implies
that there is a point like sphaleron solution. By the same logic, the
't Hooft-Polyakov monopole implies the existence of a string like sphaleron
solution in gauge theories in the Coulomb phase. We discuss the field
theory construction of the string and its supergravity dual.

\subsection{Field theory description}

We first study the unstable string in the $SU(2)$ gauge theory broken to
$U(1)$ by an adjoint Higgs. The details of the construction, the relevant
non-contractible loop in configuration space and the unstable string sitting
at the top of the loop can be found in \cite{sc,Kleihaus:1999jh}. Those
papers considered the theory in three dimensions, where the monopole is an
instanton and the sphaleron is a particle. We are interested in uplifting
this to four dimensions. We shall not review the explicit construction but
rather deduce the relevant properties from general arguments.

The monopole solution \cite{'tHooft:1974qc} yields a radial $U(1)$
magnetic field,\footnote{
We remind the reader that the $U(1)$ components of the $SU(2)$
is defined with respect to the Higgs field, $F_{\mu\nu}=F_{\mu\nu}^a W^a$.}
\bel{j1}
F_{ij}=-\frac{1}{er^3}\epsilon_{ijk}x_k,\,.
\ee\
To construct the non-contractible loop associated with this solution we
have to consider configurations which are invariant under translation in one
direction, say $x_3$. Then we replace the coordinate with a parameter
in configuration space $x_3\to\tan\alpha$.
This is pictured in \figur{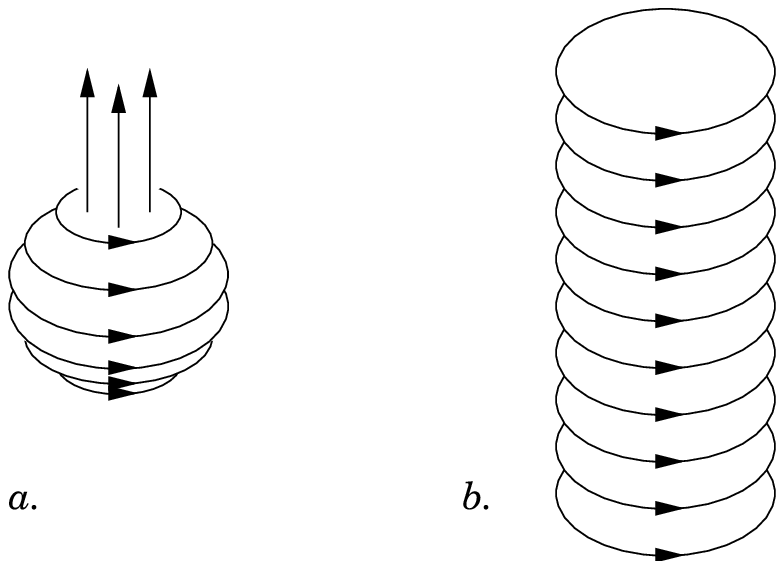}{
a. The 't Hooft-Polyakov monopole.
b. The sphaleron string is very similar to cutting the monopole in the middle
and smearing it in the $x_3$ direction. The width of the string is of order
$1/W$, where there is a non-trivial $SU(2)$ flux.}
Note that to get configurations which are independent
of the $x_3$ coordinate one has to perform an $\alpha$ dependent gauge
transformation. This does not change the topology of the loop, but it does
change the action. Therefore one cannot simply replace $x_3$ with
$\tan\alpha$ in the solution.

After the gauge transformation, the sphaleron string is given by
\be
A_a=f(x)\epsilon_{ab}x_b\sigma_3\,,
\qquad
\Phi=g(x)x_a\sigma_a\,,
\ee
with $a,b=1,2$. For more details see \cite{sc,Kleihaus:1999jh}.

For $\alpha =0$ we see (from {\it fig.} 6, (\ref{j1}) or
\cite{sc,Kleihaus:1999jh}) that there is a solution localized in the
$x^1,x^2$ plane with no magnetic flux in the plane. Thus we have
an unstable string solution (stretched along the $x_3$ direction). The string
does not carry gauge invariant $U(1)$ flux, but it does carry $SU(2)$
magnetic flux in the $x^3$ direction. Dimensional analysis implies that the
tension of such a string is
\be
T\sim\frac{W^2}{g_{YM}^2},
\eel{ten}
where $W$ is the Higgs expectation value. For $\alpha\neq 0$ there is a
$U(1)$ magnetic field and the full non-contractible loop
$-{\pi\over 2} \leq \alpha \leq {\pi\over 2}$ describes a transition which
changes the total magnetic flux of the vacuum by one unit. Note that in the
Coulomb phase this does not cost any energy as the flux expands to infinity
and we are still in the vacuum.

Put differently,  as one starts from the vacuum, $\alpha =-{\pi\over 2}$ and
goes around the non-contractible loop through the sphaleron, $\alpha =0$ back
to the vacuum $\alpha ={\pi\over 2}$, one unit of magnetic flux is added in
the $x_3$ direction. Thus the non-contractible loop goes between vacua with
different Chern numbers.

\subsection{Supergravity description}

The $AdS$/CFT correspondence is not useful to describe $SU(2)$ broken to
$U(1)$. Instead, we describe $SU(2N)$ gauge symmetry broken to
$(U(N)\times U(N))/U(1)$ by the Higgs mechanism. The relevant supergravity
background is
\cite{Maldacena:1998re}
\bel{multi}
{ds^2\over\al}={1\over
\sqrt{4\pi gN\left({1\over \vec U^4}+{1\over|\vec{U}-\vec{W}|^4}\right)}}
dx^2_{||}
+\sqrt{4\pi gN\left({1\over \vec U^4}+{1\over|\vec{U}-\vec{W}|^4}\right)}\,
d\vec{U}^2,
\ee
where $\vec{W}$ is the vector that represents the Higgs expectation value.

Since the dual of the monopole is a D1-brane in the $U$ direction and since
the sphaleron associated with the D1-brane charge is the unstable D2-brane
\cite{harvey} it is natural to suspect that the dual of the unstable string
is a D2-brane along the $x_0, x_3$ and $U$ directions. However, unlike in
$\IR^{10}$, where the boundary conditions are set at infinity, there is
nothing holding the D2-brane to the horizon. One can easily see that such
a D2-brane  will not solve the equations of motion with free boundary
conditions. Therefore, the unstable D2-brane cannot be the dual of the
unstable gauge theory string.

To resolve this puzzle we should find another object. From the discussion in
Section 4, the D2-brane can carry half a unit of D1-brane charge at each end.
Another configuration with the same charge is a D1-brane (in the $x_0,~x_3$
directions). To preserve the symmetry between $\vec U=0$ and
$\vec U=\vec W$, the D1-brane should sit precisely at the center
$\vec U=\vec W/2$. This is shown in
\figur{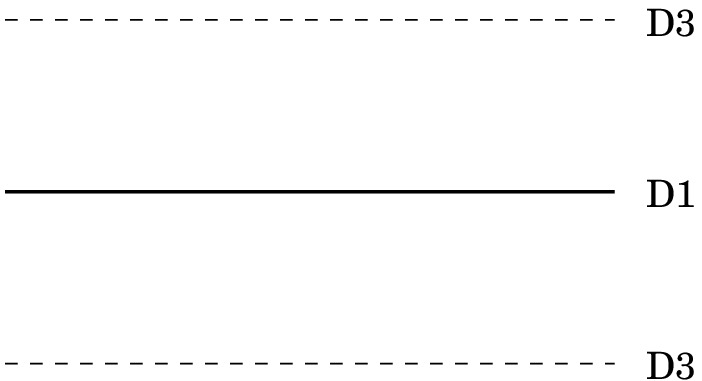}{
A D1-brane (solid line) right in between two $AdS$-like regions in the
double-centered $AdS$ geometry.}
\footnote{Other strings in this geometry were considered in
\cite{Minahan:1998xq}.}

Indeed, suppose that we place a D1-brane  along the $x_3$ direction at some
value of $\vec U$ (we could compactify the $x^3$ direction to get a finite
mass object). The field theory tension of a such a string is calculated with
respect to the field theory coordinates and is therefore
\be
T_{D1}=\frac{\sqrt{g_{00} g_{11}}}{2\pi\al g_s}\,,
\ee
 From (\ref{multi}) we see that the tension vanishes on the branes
($\vec{U}=0$ and $\vec{U}=\vec{W}$) and that the string would like to fall
towards one of the branes.

There is one exception, the string located precisely in the middle
$\vec{U}=\vec{W}/2$. It solves the classical equation of motion, however
it is unstable. Any perturbation along such a string will eventually lead
to either $\vec{U}=0$ or $\vec{U}=\vec{W}$. This is the instability of the
string in the $AdS$ description.

The tension of such a D1-brane is
\be
T\sim{W^2\over g_{YM}^2\sqrt{\lambda}}\,.
\ee
Again, we see that because this is not a BPS configuration, the tension is not
protected as one interpolates from the weakly coupled region
(\ref{ten}).

Such a D1 string carries magnetic flux in the diagonal $U(1)$ (which
decouples from the bulk degrees of freedom), but not in the relative
$U(1)$. If it falls towards one of the collections of branes, a flux is
turned on in the relative gauge group. We see that if we start with a
string at $\vec U=0$ and move it to $\vec U=\vec W$ we go back to the
vacuum, but we changed the flux in the relative gauge group by one. This
is the topological structure of the non-contractible loop and the
configuration at the middle is a sphaleron.

One can, of course, consider the configuration with a fundamental string
along the $x_0, x_3$ direction. Such a string carries an electric flux and
has the same instability. However, on the field theory side there is no dual
electric unstable string. This is an example of the case where a sphaleron
of the strongly coupled theory does not have a weakly coupled analog.
The reason is that the BPS configuration which is supposed to guarantee its
existence is the W-boson. But unlike the monopole, the W-boson is an
elementary excitation in the weakly coupled theory, and not a classical
solution, and there is no related topological charge. This is
related to the fact that the fundamental string does not carry a charge in
K-theory.

\subsection{$1/2$ Monopole configuration}

In Sections 3 and 4 we showed that the unstable D0-brane is a meron
connecting two half D-instantons.
In this subsection we generalize the construction of merons to the
't Hooft-Polyakov monopole. Consider a D1-brane in the double center $AdS$
solution (\ref{multi}) which follows one of the trajectories indicated in
\figur{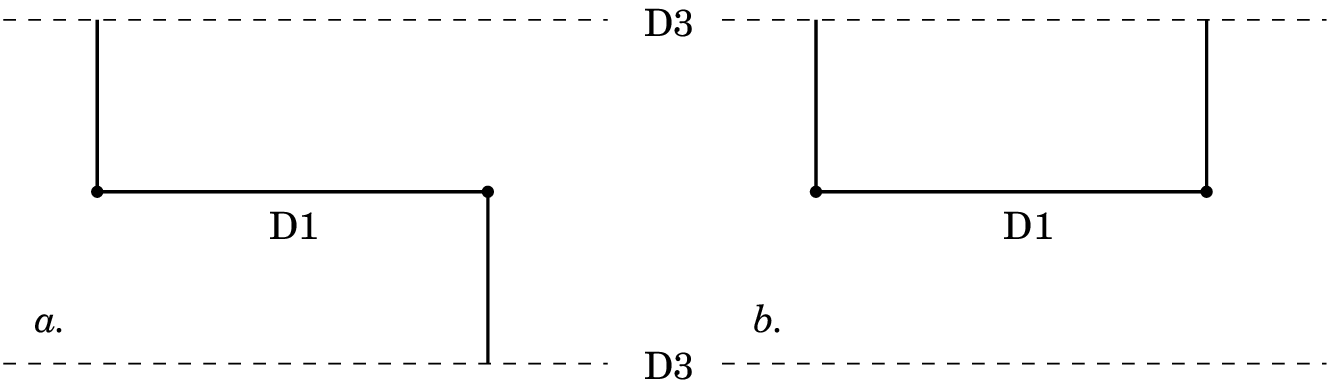}{
A D-string stretched between one horizon and the center point is half a
monopole. This configuration cannot exist on its own, but the unstable string
can connect two half-monopoles (a) or a half monopole and a half anti-monopole
(b).}
such a brane will solve the equations of motion everywhere along
the trajectory except for the two turning points.
 From the field theory side this corresponds to a monopole broken into
two half monopoles.

Notice that, unlike the meron case, the energy of the configuration is
linear with the distance between the 1/2 monopoles and hence it will not
contribute to the partition function for any value of the coupling constant.

Half a monopole seems to contradict the Dirac quantization condition.
Again there is a magnet connecting the two half monopoles. One might wonder
how this works, since we argued that this string does not carry any $U(1)$
flux. The resolution of the puzzle is simple. Recall that
the thickness of the string is $\sim 1/W$, so the string is in the
region of unbroken gauge symmetry. The flux is carried, therefore, in $SU(2)$.
See
\figur{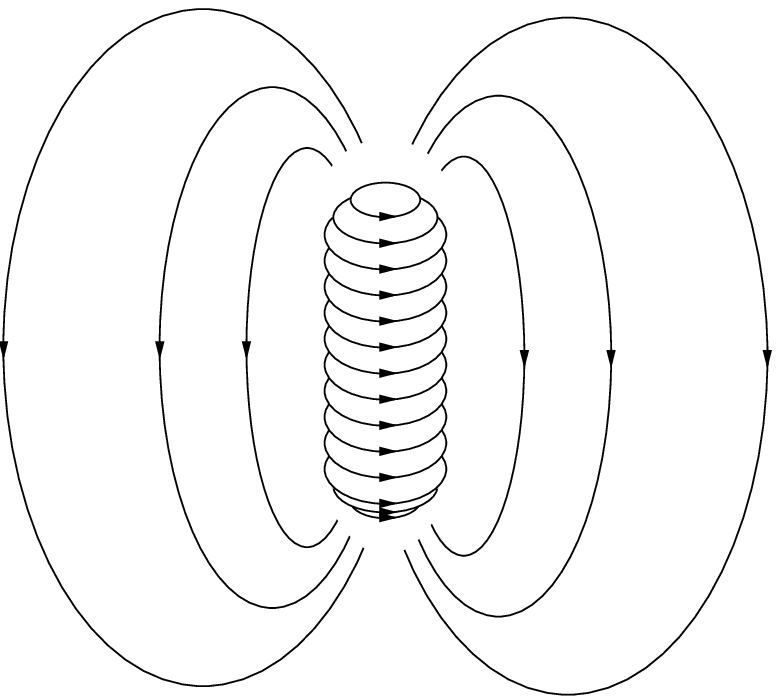}{The unstable string can end on half a monopole. Here we
draw the string with a half a monopole at one end and half an anti-monopole
at the other. From far away it looks like a $U(1)$ dipole, but near the
core, at distances of order $1/W$, a non Abelian flux is carried by the
string.}

\section*{Acknowledgments}
We are grateful to E. Gimon, P. Ho\v rava and E. Witten,
for stimulating discussions. The work on N.D. and D.J.G. is
supported by the NSF under grant No. PHY94-07194.
The work of N.I. is supported in part by the NSF under grant No. PHY97-22022.

\end{document}